\documentclass[12pt]{iopart}
\usepackage{graphicx}
\usepackage{color}
\begin{document}

\title{A quest for shear banding in ideal and non ideal colloidal rods}

\author{C. Lang$^1$, L. Porcar$^2$, H. Kriegs$^1$ and M.P. Lettinga$^1$}

\address{$^1$ ICS-3, Forschungszentrum Juelich, Juelich, Germany}
\address{$^2$ Institut Laue-Langevin, Grenoble, France}
\ead{c.lang@fz-juelich.de}
\vspace{10pt}

\begin{abstract}
We assess the possibility of shear banding of semidilute rod-like colloidal suspensions under steady shear flow very close to the isotropic-nematic spinodal, using a combination of rheology, small angle neutron scattering, and laser Doppler velocimetry. Model systems are employed which allow for a length and stiffness variation of the particles. The rheological signature reveals that these systems are strongly shear thinning at moderate shear rates. It is shown that the longest and most flexible rods undergo the strongest shear thinning and have the greatest potential to form shear bands. Although we find a small but significant gradient of the orientational order parameter throughout the gap of the shear cell, no shear banding transition is tractable in the region of intermediate shear rates. At very low shear rates, gradient banding and wall slip occur simultaneously, but the shear bands are not stable over time.
\end{abstract}

%
\vspace{2pc}
\submitto{\JPD}
%
\maketitle
%
%

\section{\label{1} Introduction}

Flow instabilities such as gradient banding can be found for many complex fluids under specific conditions. Gradient banding is the most prominent and well studied flow instability, where two regions with distinct shear rates exist along the gradient direction. Wormlike micellar systems are archetypal shear banding systems (Berret \etal 1997, Decruppe \etal 1995, Berret \etal 1998, Britton \etal 1999, Hu and Lips 2005, Miller and Rothstein 2007, Douglass \etal 2008). The combination of particle alignment, due to the particle stiffness, and the living character leads to a strong shear thinning behavior which is a prerequisite for flow instabilities. Alignment can induce flow instabilities also in non-living systems such as DNA (Boukany \etal 2008, Boukany and Wang 2010), F-actin (Kunita \etal 2012),  block co-polymer wormlike micelles (Lonetti \etal 2008) and Xanthan (Tang \etal 2018), while the formation of shear bands in flexible polymeric systems is still under debate (Tapadia and Wang 2006, Tapadia \etal 2006, Hu \etal 2007, Boukany and Wang 2009 (a) and (b)). On the other side of the material spectrum, so for quasi-ideal rod-like systems, vorticity banding under steady shear flow has been reported within the paranematic-nematic biphasic gap (Kang \etal 2006) and gradient as well as vorticity banding have been observed in the glassy state of highly charged rods (Dhont \etal 2017). Theoretical studies have shown the possibility of flow instabilities in the isotropic phase of colloidal rods sufficiently close to the isotropic-nematric (IN) phase transition (Olmsted 1999, Olmsted and Lu 1999 (a) and (b), Dhont and Briels 2003, Olmsted 2008). Indeed, rod-like colloids in the semidilute concentration regime are known to undergo strong shear thinning and concomitant biaxial orientational ordering along the flow direction when subjected to simple shear flow (Lang \etal 2016). The slope of the viscosity curve in the intermediate shear rate regime, thereby, gets steeper with increasing concentration. Notwithstanding this shear thinning behavior, there is so far no experimental evidence of gradient banding of colloidal rods in the vicinity of the nonequilibrium critical point.\par
The aim of this paper is to clarify whether the shear thinning very close to the IN transition can be strong enough to induce a shear banding instability, depending on the morphology of the rods. We explore the effect of length and flexibility, since it is known that both, rather stiff wormlike micellar systems (Lonetti \etal 2008), as well as long, semi-flexible polymers (Tang \etal 2018) can exhibit shear banding in this concentration regime due to the alignment of particles alone.\par
The systems we use are composed of rod-like bacteriophages with different morphologies, varying in length and flexibility. We bring all systems sufficiently close to the IN transition as here shear thinning is strongest, following the existing theoretical predictions. In this paper, we first use rheology to show in which shear rate region gradient banding is to be expected. We compare the slope of the flow curve in the region of lowest shear thinning coefficient $m_{fc}=d\Sigma_{21}/d\dot\gamma$, with $\Sigma_{21}$ the shear stress and $\dot\gamma$ the shear rate, to the shear thinning coefficient $m$ of the corresponding velocity profile (Tang \etal 2018). In this way, together with the form of the velocity profiles and the local order parameter $\langle P_2\rangle$, we can identify regions of gradient banding. Second, we use shear flow combined with laser Doppler velocimetry (flow-LDV) in order to investigate the velocity profiles of our systems throughout the gap. Thirdly, we use shear flow combined with small angle neutron scattering (flow-SANS) in the flow-gradient (1-2) plane in order to determine the orientational ordering throughout the gap, which can be related to the local velocity.\par
We end the paper with a discussion, linking the observations at all length scales, reaching from the microscopic rod orientation, to the mesoscopic local velocity and to the macroscopic rheology.

\section{Experiments, Materials and Methods}
\label{2}

\subsection{Experiments}
\label{2.1}

\begin{figure}[h!]
\begin{center}
\includegraphics[scale=0.7]{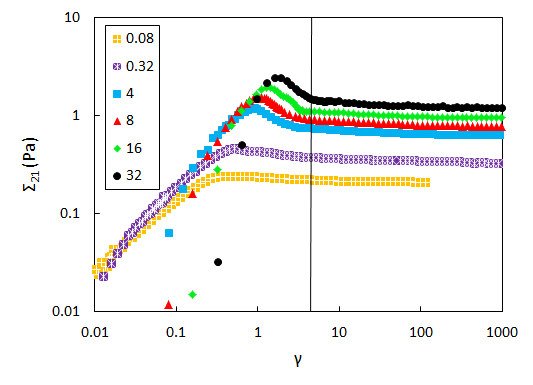}
\caption{Stress as a function of strain for different shear rates, given in units of $s^{-1}$. The line indicates the required strain of 5 strain units for reaching a steady state.}
\label{fig0}
\end{center}
\end{figure}

The rheological protocol for characterization of all samples in table~\ref{tab1} was step rate tests, where the stress is measured as a function of time under constant shear. We measured at shear rates in the range of $\dot{\gamma}\in[0.005,1000]$ $s^{-1}$ and averaged the long time-tale to determine the stress in the steady state, see figure~\ref{fig0}.

The flow-SANS study was conducted at the Institut Laue-Langevin in Grenoble, France, using their home-built shear cell in the D22 large dynamic range small angle diffractometer in order to resolve orientational ordering in the velocity-gradient plane (Liberatore \etal 2006, Helgeson \etal 2009(a) and (b)) \footnote{DOI: 10.5291/ILL-DATA.9-10-1475}.\par
All measurements were performed at 25~$^{o}$C with a detector distance of 5.6~m, a wavelength of 0.6$\pm$0.1~nm and an aperture size of 3x0.15~mm. With a total gap size of 1.35~mm, we can identify the orientational ordering at a maximum of 8 points along the gap. We use a q-range between 3.2x10$^{-2}$ and 4.6x10$^{-2}$~\AA$^{-1}$~ for the calculation of the orientational distribution function such as in earlier studies (Lang \etal 2016). For extreme low shear measurements, the shear cell has been equipped with a state of the art brushless ec-motor, providing smooth operating conditions, and a gear box transducer.\par
\begin{figure}[h!]
\begin{center}
\includegraphics[scale=0.7]{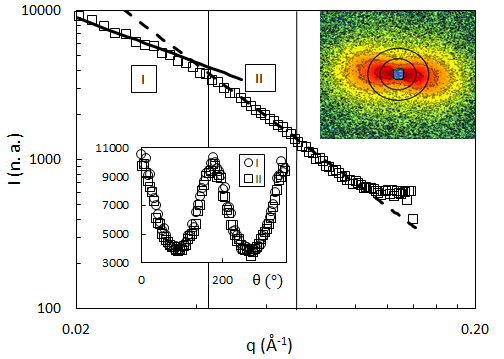}
\caption{Porod plot for Pf1. The solid and dashed lines indicate regions of $q^{-1}$ and $q^{-2}$. The vertical lines separate $q$-ranges for the analysis. Lower Inset: Comparison of the azimuthal intensity profiles for the indicated $q$-ranges. Higher Inset: 2 D scattering pattern. The black circles mark the outer borders of regions I and II.}
\label{fig1}
\end{center}
\end{figure}

Figure~\ref{fig1} shows an exemplaric intensity versus wavevector plot for fdY21M at the concentration given in table~\ref{tab1}. The full and dashed lines indicate regions of slope $q^{-1}$ corresponding to the rod-like nature of the particle and $q^{-2}$ representing random coils. Two regions, I and II, are indicated for which the wavevector averaged intensity is plotted against the azimuthal angle $\theta$ in the inset. These profiles show that the range in which the azimuthal intensity profile is obtained has no effect on the outcome of the order parameter. By fitting the azimuthal scattering intensity profiles in the given $q$-range with a previously outlined function (Lang \etal 2016) we are able to calculate the planar projection of the largest eigenvalue $\lambda_1(\theta)$ of the orientational ordering tensor in the flow-gradient plane of our experimental reference frame. The largest eigenvalue is directly proportional to the orientational order parameter $\langle P_2\rangle=(3\lambda_1-1)/2$.

Along these lines, we characterized the orientational ordering in terms of $\langle P_2(\theta)\rangle$ within the shear rate set $\dot{\gamma}\in[0.0025,64]$ $s^{-1}$ and at 5 to 8 gap positions for each shear rate. One scan of the whole gap at the usual flux at D22 with the given aperture takes 5 minutes. We carried out the same measurements for all used bacteriophages, but only the data for Pf1 will be shown here.\par
We know from our rheological investigation that the strain required to bring the Pf1 particles into a steady state of orientation amounts to 5 strain units, see figure~\ref{fig0}. Although the flow-SANS setup is not generally able to perform measurements of transient processes, for this material even at a time resolution of 5~min, time dependent effects can be studied at low shear rates, as at a shear rate of 5x$10^{-3}$ $s^{-1}$, a strain of 5 strain units is acquired only after 16.6 minutes.\par
All rheological measurements were performed with a strain-controlled ARES LS rheometer (TA Instruments, New Castle, PA, USA) at the Institute of Complex Systems 3 in Juelich, Germany. In all setups, the same Couette geometry was used. The Couette cell has an inner diameter of 32~mm with a gap size of 1~mm.\par
The laser Doppler velocimetry study was carried out at the Institute of Complex Systems 3 in Juelich Germany, where a glass Couette cell with an inner diameter of 45~mm and 1.5~mm gap size is mounted into a heterodyne dynamic light scattering setup, which is decribed elsewhere (Delgado \etal 2009).\par
The velocity profiles are normalized by the velocity of the moving inner cylinder $v_0$ and by the total gap width $y_0$. Gap scans were performed for all materials given in table~\ref{tab1}. We scanned 20 positions along the gap, measuring 5 times per point in our steady state experiments and averaging the results of the five measurements. Each point in the gap was measured for 60~s to get a stable correlation function. In the case of time dependent experiments, the measurement time per point was decreased to 3~s. This amounts to 60~s for the whole gap scan which is roughly a factor five better time resolution than in the SANS experiment. In order to minimize experimental errors, the number of repetitions of each time dependent measurement was increased by a factor 3 compared to the steady state experiment.
 
\subsection{Materials}
\label{2.2}

Three rod-like bacteriophages were grown inside their host-bacteria in Luria-Bertrani broth, following standard biological protocols (Sambrook \etal 1989). FdY21M virus is a stiff mutant of wild-type fd virus, which grows in E.-coli. This is also the case for M13k07, which is a longer more flexible mutant of wild type M13 virus. Pf1 virus is a very long and flexible Pseudomonas Aeruginosa phage which we purchased from Asla Biotech, Riga, Latvia.\par
After purification by ultra-centrifugation, the different bacteriophages were suspended in a suspending liquid holding 20 mM of Trizma base as well as 90 mM NaCl. This ensures longterm stability of the systems (at least several months) and leads to a strong binding of ions to the virus shell, resulting in a Debye length of roughly 2~nm, and thus screened Coulomb interactions between the initially negatively charged rods. For the flow-SANS measurements, the suspending liquid was Deuterium Dioxide. For the laser Doppler velocimetry measurements as well as the corresponding rheometric tests, the suspending liquid was deionized water. All bacteriophages were suspended in their buffer to reach an initial concentration of $c>c_{IN}$ and subsequently carefully diluted close to the lower IN spinodal point. The materials together with their lengths, persistence lengths and their concentrations are listed in table~\ref{tab1}. The particle polydispersity is negligible, so the lengths could be measured using atomic force microscopy. The persistence lengths were measured for fd wild-type and fdY21M using flourescence microscopy (Barry \etal 2009). Since the capsid structures of M13k07 and Pf1 are similar to that of fd wild-type, all of them have roughly the same persistence length.

\begin{table}[h]
\caption{Materials}
\smallskip{}
\centering{}
\begin{tabular}{||c|c|c|c|c||}
\hline
\textbf{material} & $L$ & $L_p$  & $c$ & $c_{IN}$ \\
 & $\mu$m & $\mu$m & mg/ml & mg/ml \\
\hline
fdY21M & 0.91 & 9.9$\pm$1.6 & 15.3 & 15.4 \\
M13k07 & 1.2 & 2.8$\pm$0.7 & 18.4 & 18.8 \\
Pf1 & 2.1 & 2.8$\pm$0.7 & 11.6 & 12 \\
\hline
\end{tabular}
\label{tab1}
\end{table}

\section{Results}
\label{3}
\begin{figure}[h!]
\begin{center}
\includegraphics[scale=0.6]{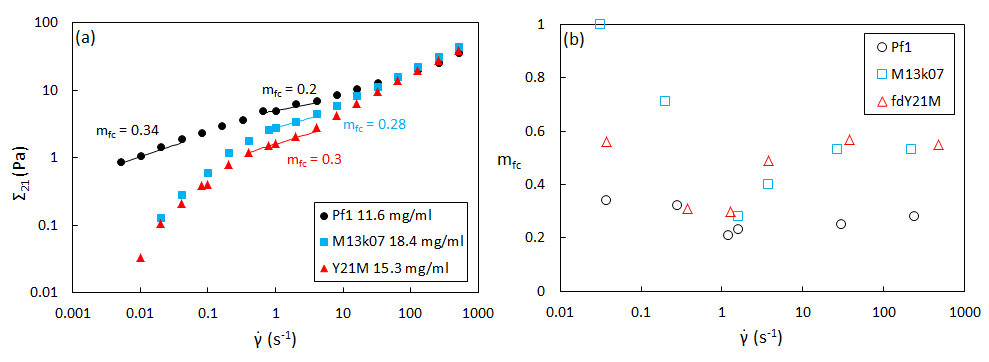}
\caption{(a) Shear stress as a function of shear rate for all viruses. The lines are fits to the data points in the indicated regions, the corresponding shear thinning coefficients are given. (b) Shear thinning coefficidents for different shear rate regimes of all systems.}
\label{fig2}
\end{center}
\end{figure}

Shear band formation is most likely to occur in the shear rate region where the observable shear thinning is strongest. In order to find the range of maximum shear thinning, we measured the flow curves of all materials given in table~\ref{tab1}, see figure~\ref{fig2}. As indicated in figure~\ref{fig2}(a), the region of the flow curve which is plateau-like is located around a shear rate of 1~$s^{-1}$ for the ideal virus fdY21M and slightly higher shear rates for the two longer viruses. We can make this observation quantitative by fitting a power law of the form $\Sigma_{21}=K\dot\gamma^{m_{fc}}$ to the flowcurve, selecting certain shear rate ranges. In this region, we find the minimum shear thinning coefficients $m_{fc}$, corresponding strongest shear thinning, see the lines in figure~\ref{fig2}(a) and compare to figure~\ref{fig2}(b). The viscosity ratio between the beginning and the end of this region of low slope is for all our materials $\eta(\dot\gamma_\mathrm{low})/\eta(\dot\gamma_\mathrm{high})\approx3.5$.  All flow curves overlap at high shear rates, where the orientational ordering of viruses saturates. At lower shear rates, the two curves of fdY21M and M13k07 have a similar slope, while that of Pf1 is less steep. The slope of the curves in this low shear rate regime corresponds to the low shear viscosity. Note that this low shear viscosity should not be confused with the zero shear viscosity despite the fact that it shows a similar trend, because the zero shear viscosity is not measurable for the given systems, as has been shown in a previous paper (Lang \etal 2016).

\noindent{}Figure~\ref{fig2} shows that the shear thinning coefficient in this region decreases with increasing length of the particles. Therefore, the longest viruses Pf1 are the most likely to undergo gradient banding.\par
\begin{figure}[h!]
\begin{center}
\includegraphics[scale=0.6]{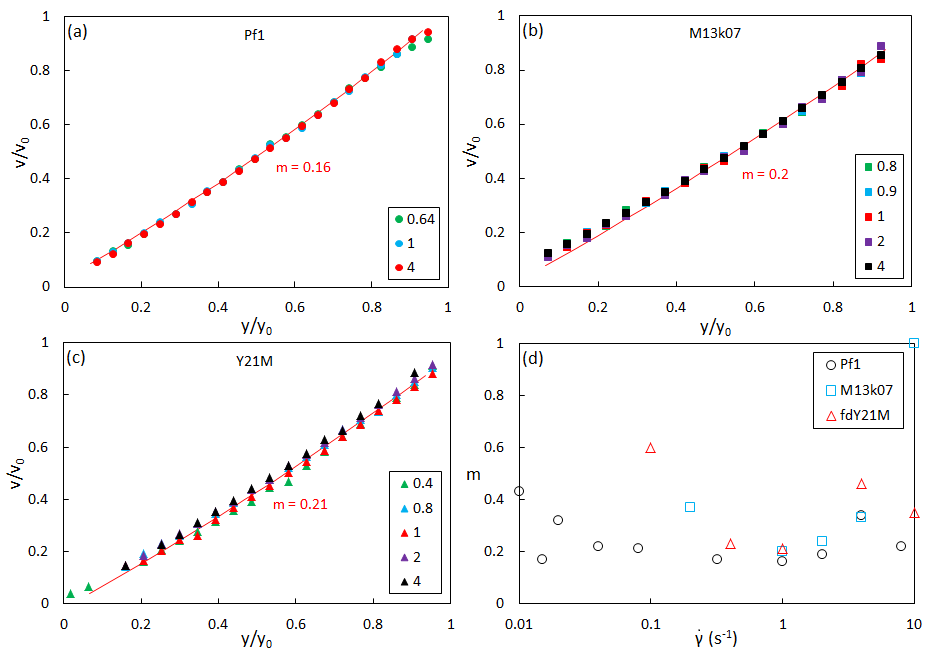}
\caption{Relative velocity versus relative gap position for (a) Pf1, (b) M13k07 and (c) fdY21M. The lines are fits to the data points with the respective colour. The legend shows the shear rate values of the different experiments.}
\label{fig3}
\end{center}
\end{figure}

In order to directly visualize the flow gradients, we use laser Doppler velocimetry and evaluate the velocity of the sheared samples along the gradient direction. Figure~\ref{fig3} shows the velocity profiles measured in the strongest shear thinning regions of each system. It is obvious that none of the velocity profiles display larger nonlinearities or shear bands. The lines in figure~\ref{fig3}(a), (b) and (c) show curve fits to the measurement data. As the velocity profile $v(y)$ in a Couette cell for for a shear thinning system is per defintion nonlinear, we use the following relation (Tang \etal 2018):
$$v(y)=\dot\gamma r_i\frac{(r_o-y)^{1-2/m}-r_o^{1-2/m}}{r_i^{1-2/m}-r_i^{1-2/m}}~~,$$
with inner radius of the Couette cell $r_i$ and outer radius $r_o$ located along the gradient direction $y$. We find the shear thinning coefficient $m$ by employing the least squares method. The shear thinning coefficients are shown for selected shear rates in figure~\ref{fig3}(d).

The shear thinning coefficients obtained from the velocity profiles show the same trend as those obtained from the flow curves, reaching minimum values around $\dot\gamma=1~s^{-1}$ for all systems. As for the shear thinning coefficients obtained from the flow curves, also $m$ decreases with increasing length and flexibility of the constituent particles.\par
\begin{figure}[h!]
\begin{center}
\includegraphics[scale=0.6]{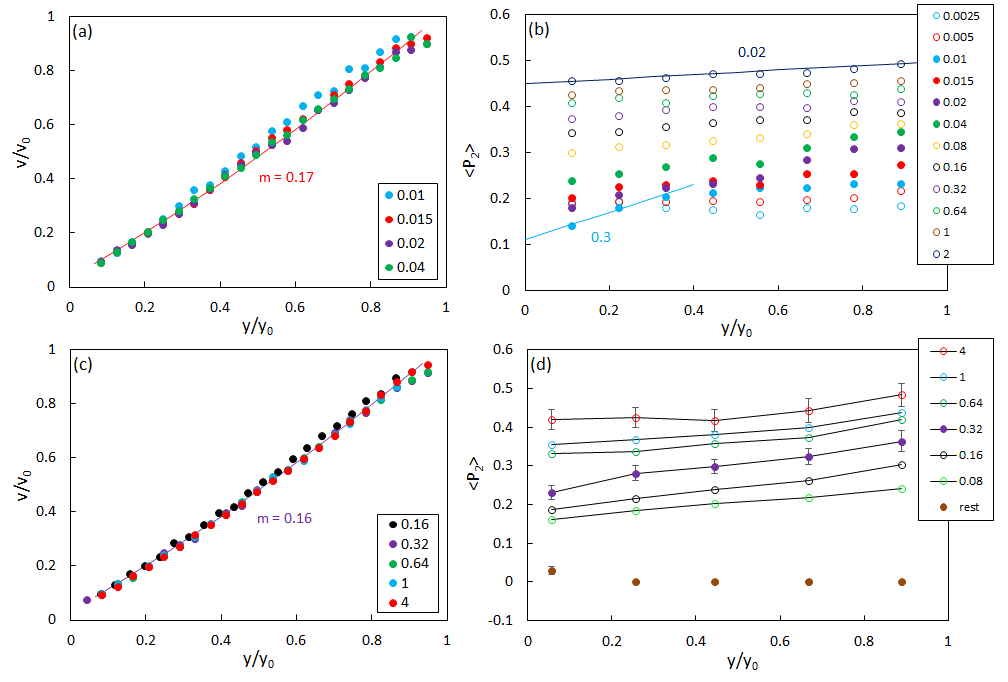}
\caption{Relative velocity versus relative gap position for Pf1 in (a) the low shear rate regime and (c) at intermediate shear rates. The lines indicate fits to the data at 0.015 and 0.32~$s^{-1}$. Order parameter versus relative gap position in (b) the low shear rate regime and (d) at intermediate shear rates, the lines indicate the gradient of $\langle P_2\rangle$.}
\label{fig4}
\end{center}
\end{figure}

Interestingly, the long rod Pf1 is the only species showing another shear rate region of minimum shear thinning coefficient located at very low shear rates, see figure~\ref{fig3}(d). This region is also indicated by the low shear rate line in figure~\ref{fig2}(a). Figure~\ref{fig4}(a) shows the velocity profiles of Pf1 in this low shear rate region. As in the intermediate shear rate regime, the velocity profiles do not display any gradient bands, but the shear thinning coefficients seem to vary more strongly from one shear rate to the other, as compared to higher shear rates, see figure~\ref{fig3}(d).

The orientational ordering corresponding to the local shear rate at different positions along the gradient axis was measured using flow-SANS. Since the local shear thinning coefficient of the long, flexible species Pf1 at very low and intermediate shear rates reaches by far the lowest values, we study the local alignment of these molecules, see figures~\ref{fig4}(b) and (d). Even in the absence of shear, we find a significant ordering of the rods close to the wall, $y/y_0\to0$, see yellow symbols in figure~\ref{fig4}(d). We attribute this to wall-anchoring of the rods, which entropically prefer to order perpendicular to the surface normal vector of the wall, leading to wall induced nematization (Dijkstra \etal 2001, Klop \etal 2018).
 Surprisingly, the wall anchoring in this system seems to persist for a distance of at least 30 particle lengths. While most of the alignment curves have a small gradient $d\langle P_2\rangle/dy\approx0.02$, we identify a few shear rates at which the local alignment drops significantly towards the standing wall, $\lim_{y/y_0\to0}d\langle P_2\rangle/dy\approx 0.3$. In the intermediate shear rate regime we find a significant drop of the order parameter closer to the standing wall for a shear rate of 0.32~$s^{-1}$, see solid symbols in figure~\ref{fig4}(d). In the very low shear rate regime, the same behavior is found for shear rates between $\dot\gamma\in[0.01,0.04]$~$s^{-1}$, see figure~\ref{fig4}(b). A comparison to the velocity profiles at exactly these shear rates indicates that despite this observation, no gradient shear banding is found in the steady state.\par
\begin{figure}[h!]
\begin{center}
\includegraphics[scale=0.6]{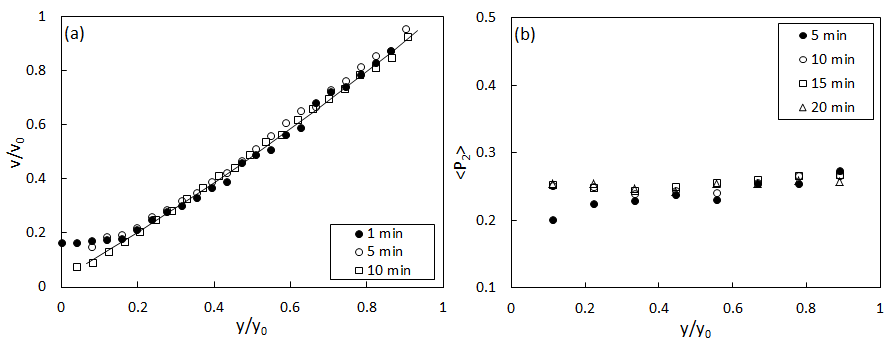}
\caption{Gap scan of Pf1 at a shear rate of 0.015~$s^{-1}$ in (a) the velocity and (b) the orientational ordering. The line indicates the velocity profile for a shear thinning coefficient of 0.17.}
\label{fig5}
\end{center}
\end{figure}

Due to the marked differences between the orientational ordering close to the standing wall for the shear rates 0.005 and 0.01 $s^{-1}$, and given the extremely low shear rates, it should be tested if there is an effect of the acquired strain units, and thus a time dependence of the velocity profile as well as the order parameter in the low shear rate regime. Figure~\ref{fig5} shows a comparison between the time dependent and steady state curves for a shear rate of 0.015~$s^{-1}$.  

Indeed, we can identify a significant difference between the short time and steady state curves in both figures~\ref{fig5}(a) and (b), indicating gradient shear banding including wall slip within the first 5 minutes of the experiment, which vanishes with increasing time. Both, the gradient in order parameter, as well as the significantly bent velocity profile disappear after about 10 minutes, amounting to roughly to 9 strain units, which is almost twice the strain necessary for reaching the steady state (see figure~\ref{fig0}). No further time dependent changes of the order parameter could be found.

\section{Discussion}
\label{5}

The goal of this paper was to investigate if ideal colloidal rods are prone to undergo a flow instability in the form of gradient shear bands and if this can be tuned by the rod morphology. Here we call a rod-like system ideal if the aspect ratio is larger than 100 and the persistence length is significantly larger than the length of the particle, according to the equilibrium theory for the isotropic-nematic phase transition by Onsager (Onsager 1949). Indeed, Barry \etal showed that fdY21M has the phase transition exactly at the predicted volume fraction. It has also been shown that these rods follow the non-equilibrium theory for sheared nematics, based on theory of Olmsted 1999 and Dhont and Briels 2003, see Lettinga \etal 2005. Around the I-N phase transition, the collective rotational diffusion goes to zero, which makes the system highly susceptible to shear flow. In figure~\ref{fig2}, we compare the flow curves of rod-like particles with different morphologies very close to the IN transition. All of the curves display the strongest shear thinning behavior with the corresponding smallest shear thinning coefficient in the intermediate shear rate regime. The shear thinning coefficient in this region decreases with increasing rod length and flexibility. However, in comparison to other systems in literature, the shear thinning coefficients found here seem to be on the edge for the possibility of gradient banding. Only the longest system, Pf1 with a length of 2.1~$\mu$m, displays a shear thinning coefficient that is close to values found for other shear banding systems, i. e. $m_{fc}\approx0.2$, such as in semi-flexible Xanthan gum (Tang \etal 2018). For wormlike micellar systems $m_{fc}$ often becomes essentially 0 or even negative (Berret \etal 1997, Gaudino \etal 2017).\par
Figures~\ref{fig3}(a), (b) and (c) show the steady state velocity profiles of all our systems in the interesting shear rate range and beyond. None of the systems show any peculiarities nor shear bands separated by a sharp interface, not even Pf1. This, however, does not mean that no flow instability is present. If the shear thinning coefficient $m$, measured from the velocity profile, is much smaller than $m_{fc}$, obtained from the flowcurve, then this is an indication that the flow is unstable and possibly shear banding with a very broad interface. The lack of shear band formation is confirmed by the flow-SANS scans of the gap, which show that there is no appreciable differences in the orientation throughout the gap in the region where the shear thinning is highest.\par 
It is remarkable that in the past the origin of shear banding has been attributed to proximity of the IN transition for systems that are far from being ideal rods, such as surfactant wormlike micelles (Decruppe \etal  1995, Berret \etal 1998, Britton \etal 1999), whereas systems that could be considered ideal rods do not display shear banding, despite of the fact that they are perfectly stiff, slender and monodisperse, such as fdY21M. That shear band formation can indeed be intimately related with the proximity of the IN transition has been shown for surfactants (Helgeson \etal 2009(b)) and block-copolymer wormlike micelles (Lonetti \etal 2008). Especially the latter system bares similarities with the systems under study as it does not have scission as an extra relaxation mechanism, except that it is polydisperse in length with a tail in the distribution of worms with a length of several micrometers. One possible reason is that vorticity bands can form when shearing a system of rods, which is in the IN phase coexisting region at rest (see Kang \etal 2006). Here, we choose to go up to the IN binodal point, but not past it. However, we know also from simulations and experiments that the nonequilibrium binodal line has a positive slope in the shear rate versus concentration diagram, such that the system will stay thermodynamically stable when increasing the shear rate (Ripoll \etal 2008). Though we came as close as possible to the nonequilibrium critical point, as predicted by Olmsted 1999 and Dhont and Briels 2003, we also know from the latter theoretical studies that a shear banding transition becomes likely only very close to the IN transition $c/c_{IN}\approx0.96$ and over a very narrow shear rate range. Hence, possibly shear banding for ideal rods takes only place in a region that is not accessible as other instabilities kick in.\par
The fact that Pf1 is closest to display a flow instability suggests that also the absolute length of the rods plays an important role. Indeed, when comparing the rods under study with stiff systems that do show shear banding, such as Xanthan (Tang \etal 2018), Ehut (van der Gucht \etal 2006), DNA (Boukany \etal 2009), PB-PEO (Lonetti \etal 2008)and F-actin (Kunita \etal 2012), then it is noteworthy that all these systems contain particles with a length which is multiple times the length of Pf1. Moreover, it is not clear for any of these systems whether they are close to the biphasic regime, which is most likely due to the comparatively high particle length slowing down all kinetics, including phase separation. Hence the systems are likely metastable at the moment they are subjected to shear flow. We conclude that for shear banding to occur in rod-like systems, the rods need to be long enough. One reason might be that a higher particle length will enhance the contrast between bands in terms of the viscosity and orientation, which is needed to sustain an interface between bands. We note again that the Pf1 virus, although the longest system in this study, is significantly shorter than for example Xanthan gum (Tang \etal 2018) or wormlike micelles (Helgeson \etal 209(b)), for both of which the low to high shear rate viscosity ratio in the region of lowest shear thinning parameter is about a factor 10 higher than in our system. The largest gradient in orientational order observed for our system $d\langle P_2\rangle/dy\approx 0.3$ is even more than a factor 100 lower than in wormlike micelles, where $d\langle P_2\rangle/dy\approx 35$ (Helgeson \etal 2009(b)). Accoringly, also the ratio in viscosity between the high and low shear band is much lower. For Pf1, we estimate that this ratio is given by $\eta(\dot\gamma_\mathrm{low})/\eta(\dot\gamma_\mathrm{high})\approx3.5$, while values for other systems with proven shear banding are much higher: Xanthan $\eta(\dot\gamma_\mathrm{low})/\eta(\dot\gamma_\mathrm{high})\approx20$ (Tang \etal 2018), wormlike micelles $\eta(\dot\gamma_\mathrm{low})/\eta(\dot\gamma_\mathrm{high})\approx40$ (Helgeson \etal 2009(b)), or DNA $\eta(\dot\gamma_\mathrm{low})/\eta(\dot\gamma_\mathrm{high})\approx1500$ (Boukany and Wang 2009(b)). Despite its high rigidity in comparison with Xanthan and wormlike micelles, it should be mentioned that Pf1 cannot be considered an ideal rod, since by the criterium of Odijk 1983, it can bend significantly, thus it probably forms a highly entangled state. Given a certain theoretical tube diameter $D$ for a system of rods with number density $\nu$, the flexibility of the particles plays a role if $D<L<\left(D^2L_p\right)^\frac{1}{3}$. The occurring tube diameter can be approximated by the mesh size of the particle network $D\sim\nu L^2$.\par
While we do not observe bands in the expected shear rate region, we do observe that the velocity profiles of Pf1 display a stronger variety of curvature in the very low shear rates region, yielding $m=0.17<m_{fc}=0.34$. More interestingly, the order parameter for shear rates $\dot\gamma\in[0.01,0.04]$ decreases significantly towards the standing wall as compared to that of  shear rates $\dot\gamma\in[0.0025,0.005]$. This behavior is comparable to that of phase separating wormlike micellles (Liberatore\etal 2006), although it occurs in a very different shear rate regime. Focusing on $\dot\gamma=0.15$, we even observe a shear rate band, both in the orientational ordering and velocity, which is however transient, see figure~\ref{fig5}, as it relaxes over a duration of 10~min, corresponding to an uptake of 9 strain units. Thus, the initially sharp interface between gradient bands cannot be sustained over time and decomposes upon the uptake of enough strain into a steady state $m<m_{fc}$ with a comparatively high curvature of the flowcurve. This behavior is opposite of what has been shown for DNA (Boukany and Wang 2009(b)), where an increase in acquired strain helps shear band formation. Judging from the velocity profiles, the disappearance of the shear bands comes along with the disappearance of wall-slip, as observed for wormlike micelles (Britton \etal 1999, Hu and Lips 2005).  In contrast to polymers, these system mostly slip at the standing wall, which seems to be the case for Pf1 as well.\par
The transient behavior might also be related with the fact that we do observe wall induced alignment due to entropic wall anchoring, see figure~\ref{fig4}(d). Already very low applied shear rates $\dot\gamma\in[0.0025,0.005]$ orient all rods thoughout the gap significantly more than the initially wall anchored state, compare figures~\ref{fig4}(b) and (c). Since the observed time dependent gradient banding occurs at higher shear rates, we can rule out the possibility that this gradient banding is caused by wall-anchoring of the rods.\par

\section{Conclusions}
\label{6}
Using rod-like bacteriophages with varying morphology, we are able to describe the rheological behavior of rod-like suspensions close to the IN transition and assess their potential to undergo flow instabilities. With increasing length and flexibility of the particles, these systems show stronger shear thinning reaching a point where comparable polymeric systems undergo gradient banding. No bands are formed in the intermediate shear rates regimes, but at very low shear rates, time dependent gradient banding is found. The low shear rate gradient bands occur simultaneously with wall-slip and possess a sharp interface between the two bands. This interface broadens with time until the bands effectively disappear after the acquired strain is large enough. In combination with findings on wormlike micelles and polymers from the literature, we conclude that the interplay between particle entanglement and orientational ordering suppresses shear banding of rod-like systems in the steady state.

\ack
This research is funded by the European Union within the Horizon 2020 project under the DiStruc Marie Sk\l{}odowska Curie innovative training network; Grant Agreement No. 641839. 

 \section*{References}
\begin{harvard}

\item[] Barry E., Beller D. and Dogic Z. A model liquid crystalline system based on rodlike viruses with variable chirality and persistence length. 2009 {\textit Soft Matter} {\bf 5}, 2563-2570
\item[] Berret J. F., Porte G. and Decruppe J. P. Inhomogeneous shear flows of wormlike micelles: A master dynamic phase diagram. 1997 {\textit Phys. Rev.} E {\bf 55}, 1668-1676
\item[] Berret J. F., Roux D. C. and Lindner P. Structure and rheology of concentrated wormlike micelles at the shear induced isotropic to nematic transition. 1998 {\textit Eur. Phys. J.} B {\bf 5}, 67-77
\item[] Boukany P. E., Hu Y. T. and Wang S. Q. Observations of wall slip and shear banding in an entangled dna solution. 2008 {\textit Macromolecules} {\bf 41}, 1644-2650
\item[] Boukany P. E. and Wang S. Q. Shear banding or not in entangled dna solutions depending on the level of entanglement. 2009 {\textit J. Rheol.} {\bf 53}, 73-83
\item[] Boukany P. E. and Wang S. Q. Exploring the transition from wall slip to bulk shear banding in well entangled dna solutions. 2009 {\textit Soft Matter} {\bf 5}, 780-789
\item[] Boukany P. E. and Wang S. Q. Shear banding or not in entangled dna solutions. 2010 {\textit Macromolecules} {\bf 43}, 6950-6952
\item[] Boukany P. E., Wang S. Q., Ravindranath S. and Lee L. J. Shear banding in entangled polymers in the micron scale gap: a confocal rheoscope study. 2015 {\textit Soft Matter} {\bf 11}, 8058-8068
\item[] Britton M. M., Mair R. W., Lambert R. K. and Callaghan P. T. Transition to shear banding in pipe and Couette flow of wormike micellar solutions. 1999 {\textit J. Rheol.} {\bf 43}, 897-909
\item[] Decruppe J. Pl. and Cressely R. and Makhloufi R. and Cappelaere E. Flow birefringence showing a shear banding structure in a ctab solution. 1995 {\textit Colloid. Polym. Sci.} {\bf 273}, 346-351
\item[] Delgado J.,  Kriegs H and Castillo  R 2009 \textit{J. Phys. Chem} B {\bf113}, 1548515494
\item[] Dhont J K G and  Briels W J 2003 \textit{Colloid. Surface.} A {\bf213},131
\item[] Dhont J. K. G., Kang, K., Kriegs, H., Danko O., Marakis J. and Vlassopoulos D. Nonuniform flow in soft colloidal rods. 2017 {\textit Phys. Rev. Fluids} {\bf 2}], 043301
\item[] Dijkstra M., van Roij R. and Evans, R. Wetting and capillary nematization of a hard-rod fluid: A simulation study. 2001 {\textit Phys. Rev.} E {\bf 63}, 051703
\item[] Douglass B. S., Colby, R. H. Madsen L. A. and Callaghan P. T. Rheo nmr of wormlike micelles formedfrom nonionic pluronic surfactants. 2008 {\textit Macromolecules} {\bf 41}, 804-814
\item[] Gaudino D.,  Pasquino R., Kriegs H., Szekely N.,  Pyckhout-Hintzen W.,  Lettinga M. P. and Girzzuti N. Effect of the salt-induced micellar microstructure on the nonlinear shear flow behavior of ionic cetylpyridinium chloride surfactant solutions. 2017 \textit{Phys. Rev. } E {\bf95}, 032603
\item[] Helgeson M. E.,  Vasquez P. A.,  Kaler E. W. and  Wagner N. J. 2009 \textit{J. Rheol} {\bf53}, 727
\item[] Helgeson M. E., Reichert, M. D., Hu Y. D. and Wagner N. J. Relating shear banding, structure, and phase behavior in wormlike micellar solutions. 2009 {\textit Soft Matter} {\bf 5}, 3858–3869
\item[] Hu Y.T. and Lips A. Kinetics and mechanism of shear banding in an entangled micellar solution. 2005 {\textit J.  Rheol.} {\bf 5}, 1001-1027
\item[] Hu Y. T., Wilen L., Philips A. and Lips A. Is the constitutive relation for entangled polymers monotonic? {\textit J. Rheol.} {\bf51}, 275-295 
\item[] Kang K., Lettinga M. P., Dogic, Z. and Dhont J. K. G. Vorticity banding in rodlike virus suspensions. 2006 {\textit Phys. Rev.} E {\bf 74}, 026307
\item[] Klop K. E., Dullens R. P. A., Lettinga M. P., Egorov S. A. and Aarts D. G. A.  L. Capillary nematisation of colloidal rods in confinement. 2018 {\textit Molecular Physics}, DOI: 10.1080/00268976.2018.1497210
\item[] Kunita I., Sato K. Tanaka Y., Takikawa, Y., Orihara H. and Nakagaki, T. Shear banding in an f-actin solution. 2012 {\textit Phys. Rev. Lett.} {\bf 109}, 248303
\item[] Lang C., Porcar L., Kohlbrecher J. and Lettinga M. P. The connection between biaxial orientationa and shear thinning for quasi-ideal rods. 2016 {\textit Polymers} {\bf 8}, 291
\item[] Lettinga M. P., Dogic Z., Wang H. and Vermant J. Flow behavior of colloidal rodlike viruses in teh nematic phase. 2005 {\textit Langmuir} {\bf 21}, 8048-8057
\item[] Liberatore M. W.,Nettesheim  F. and  Wagner N. J. Spatially resolved small-angle neutron scattering in the 1-2 plane: A study of shear-induced phase-separating wormlike micelles 2006 \textit{Phys. Rev.} E {\bf73}, 020504
\item[] Lonetti B., Kohlbrecher J., Willner L., Dhont J. K. G. and Lettinga M. P. Dynamic response of block copolymer wormlike micelles to shear flow. 2008 {\textit J. Phys.: Condens. Matter} {\bf20}, 404207
\item[] Miller E. and Rothstein J. Transient evolution of shear banding wormlike micellar solutions. 2007 {\textit J. Nonnewton. Fluid. Mech.} {\bf 143}, 15
\item[] Odijk, T. On the statistics ansd synamics of confined or entangled stiff polymers. 1983 {\textit Macromolecules} {\bf 16}, 1340-1344
\item[] Olmsted P. D. Two-state shear diagrams for complex fluids in shear flow. 1999 \textit{Curr. Opin. Colloid. In.} {\bf4}, 95
\item[] Olmsted P. D. and  Lu C.-Y. D. Phase coexistence of complex fluids in shear flow. 1999 \textit{Faraday Discuss.} {\bf112}, 183
\item[] Olmsted P. D. and  Lu C.-Y. D. Phase separation of rigid-rod suspensions in shear flow. 1999 \textit{Phys. Rev.} E {\bf60}, 4397
\item[] Olmsted P. D. Perspectives on shear banding in complex fluids. 2008 \textit{Rheol. Acta} {\bf47}, 183300
\item[] Onsager L. The effects of shape on the interaction of colloidal particles. 1949 {\textit Ann. N.Y. Acad. Sci.} {\bf 52} 627-659
\item[]Ripoll M.,  Holmquist P.,  Winkler R. G., Gompper  G.,  Dhont J. K. G. and Lettinga M. P. Attractive Colloidal Rods in Shear Flow. 2008 \textit{Phys. Rev. Lett.} {\bf101}, 168302
\item[] Sambrook J,  Fritsch E and  Russel M T 1989 \textit{Molecular Cloning: A laboratory manual} (Cold Spring Harbor Laboratory Press)
\item[] Tang, H., Kochetkova T., Kriegs H., Dhont J. K. G. and Letttinga M. P. Shear banding in entangled xanthan solutions: tunable transition from sharp to broad shear band interfaces. 2018 {\textit Soft Matter} {\bf 14}, 826-836
\item[] Tapadia P. and Wang S. Q. Direct visualization of continuous simple shear in non-newtonian polymeric fluids. 2006 {\textit Phys. Rev. Lett. } {\bf 96}, 016001
\item[] Tapadia P., Ravindranath S. and Wang S. Q. Banding in entangled polymer fluids under oscillatory shearing. 2006 {\textit Phys. Rev. Lett. } {\bf 96}, 196001
\item[] van der Gucht J., Lemmers M., Knoben W., Besseling M. A. N. and Lettinga, M. P. Multiple Shear-Banding Transitions in a Supramolecular Polymer Solution. 2006 {\textit Phys. Rev. Lett.} {\bf 97}, 108301

%
%
%
%
%
\end{harvard}

\end{document}